# Finite element 3D Modelling of Mechanical Behaviour of Mineralized Collagen Microfibrils

Abdelwahed BARKAOUI, Ridha HAMBLI
PRISME Laboratory, EA4229, University of Orleans
Polytech' Orléans, 8, Rue Léonard de Vinci 45072 Orléans, France
ridha.hambli@univ-orleans.fr
abdelwahed.barkaoui@univ-orleans.fr

**Abstract.** The aim of this work is to develop a 3D finite elements model to study the nanomechanical behaviour of mineralized collagen microfibrils, which consists of three phases, (i) collagen phase formed by five tropocollagen (TC) molecules linked together with cross links, (ii) a mineral phase (Hydroxyapatite) and (iii) impure mineral phase, and to investigate the important role of individual properties of every constituent. The mechanical and the geometrical properties (TC molecule diameter) of both tropocollagen and mineral were taken into consideration as well as cross-links, which was represented by spring elements with adjusted properties based on experimental data. In the present paper an equivalent homogenised model was developed to assess the whole microfibril mechanical properties (Young's modulus and Poisson's ratio) under varying mechanical properties of each phase. In this study both equivalent Young's modulus and Poisson's ratio which were expressed as functions of Young's modulus of each phase were obtained under tensile load with symmetric and periodic boundary conditions.

1. Introduction

Hierarchical structures in bio-composite such as bone tissue have many scales or levels, specific interactions between these levels and highly complex architecture in order to achieve its biological and mechanical functions (1). These complexity and heterogeneity of bone tissue require a multiscale modelling to understand its mechanical behaviour and its remodelling mechanism (2). In long bones, like the femur, three parts are distinguished, from the center outward: the marrow, the spongy bone and cortical bone. Human cortical bone structure consists of six structural scale levels which are the (macroscopic) cortical bone, osteonal, lamellar, fibrous, fibril and microfibril Fig. 1. In fact, microscopic analysis reveals a complex architecture that can be described as hollow cylinders juxtaposed next to each other and sealed by a matrix. The cylinders are called osteon, the holes are named Haversian canals and the matrix is the interstitial system. Further analysis shows that osteons are in fact an assembly of cylindrical strips embedded in each other and each lamella is composed of a network of collagen fibers with a helical orientation and inserted into hydroxyapatite crystals. The orientation of collagen fibers may be different between two consecutive lamellae. Every fiber is formed by a set of fibrils. Each fibril is in turn composed of microfibrils. Finally, each microfibril is a helical arrangement of five TC molecules (3).

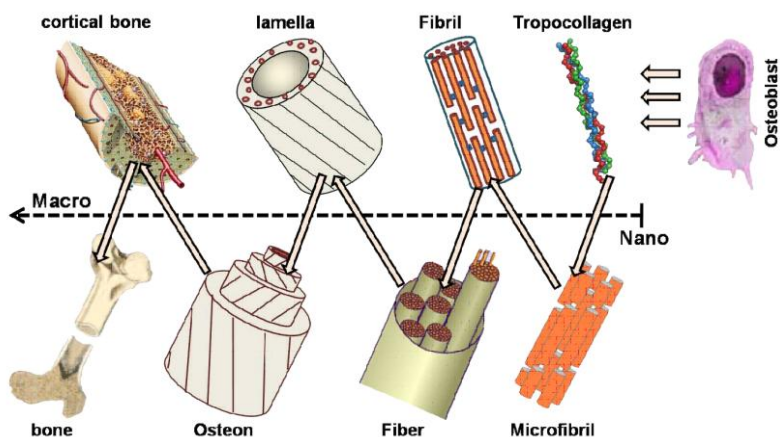

**Figure1.** The multiscale hierarchical structure of cortical bone

At ultrastructure level, mineral and collagen are arranged into higher hierarchical levels to form microfibrils, fibrils and fibers (4-6). The existence of sub-structures in collagen fibrils has been a debate for years. Recent studies suggest the presence of microfibrils in fibrils, experimental works prove that all collagen-based tissues are organized into hierarchical structures, where the lowest hierarchical level consists of triple helical collagen molecules (7-9) and the multiscale structure was defined as triple helical collagen molecules - microfibrils – fibrils - fibers. A longitudinal microfibrillar structure with a width of 4 - 8 nm was visualized in both hydrated (10) and dehydrated (11). Three-dimensional image reconstructions of 36 nm-diameter corneal collagen fibrils also showed a 4 nm repeat in a transverse section, which was related to the microfibrillar structure (12). Using X-ray diffraction culminating in an electron density map, (8) suggested the presence of right-handed super twisted microfibrillar structures in collagen fibrils.

Experimentally, single tropocollagen molecules and mechanical properties of fibrils of mineralized collagen studies have been performed by several authors (13-15). Other studies on the hydrated collagen microfibril in the small strain regime based on X-ray diffraction (16, 17) and atomic force microscopy (AFM) (9-15) were also achieved to investigate the stress-strain relation and the elastic properties of microfibrils. Further more study of the effect of collagen-mineral deformation processes (18) was realized to estimate the mechanical properties of mineralized collagen fibrils and bone tissue. Recently an atomistic model of collagen microfibril dynamic has been developed (5). This model in fact interprets in full details the mechanical behavior at the microfibril level, on other side it requires extremely computational power due to its size. In spite of all these efforts towards this scale, a 3D FE model which can represent ultrastructre of mineralized collagen that takes into consideration the TC, mineral phases as well as the cross-links are still absent.

In the current work as an attempt to be more close to the reality, a 3D FE model is proposed to study the mechanical behavior of mineralized collagen microfibril considering these aspects: (i) the dimensions and composition of the constituents (mineral, TC molecules and gross-links). (ii) Studying the effect of elastic mechanical properties of constituents on the whole microfibril structure. The proposed 3D finite element model gives the ability to study bone mechanical properties at nanoscopic scale and permit easily parametric studies to investigate the effect of mechanical and geometrical parameters (mineral density, cross-links, elastic properties of bone component,…etc) related to bone quality.

## 2. Methods and tools

In this section we discuss: (i) the structure of the microfibril and the mechanical behaviour of these elementary constituents, (ii) selection and development of 3D finite element model.

The microfibril is a helical assembly of five TC molecules (rotational symmetry of order 5), which are offset one another with apparent periodicity of 67 nm Fig. 2. This periodical length is denoted by the letter D and it's used as a primary reference scale in describing the structural levels. The helical length of a collagen molecule is 4.34 D ≈ 291 nm and the discrete gap (hole zone) is 0.66 D ≈ 44 nm ( = 35 nm in some other references) between two consecutive TC molecules type1 in a strand. These gaps in bone are the sites of nucleation for hydroxyapatite crystals (the mineral component of bone tissue) to be deposited (19, 20). Those five molecules create a cylindrical formation with a diameter 3.5–4 nm and its length is unknown. The orientation and axial arrangement of TC molecules in the microfibril have been deducted from an electron-microscopic observation showing transverse striations with a period D. The origin of this streaking was performed by a gradation in the arrangement of elements that are staggered tropocollagen with themselves at an interval D. The strength and stability during maturation of the microfibrils are achieved by the development of intermolecular cross-links (21, 22).

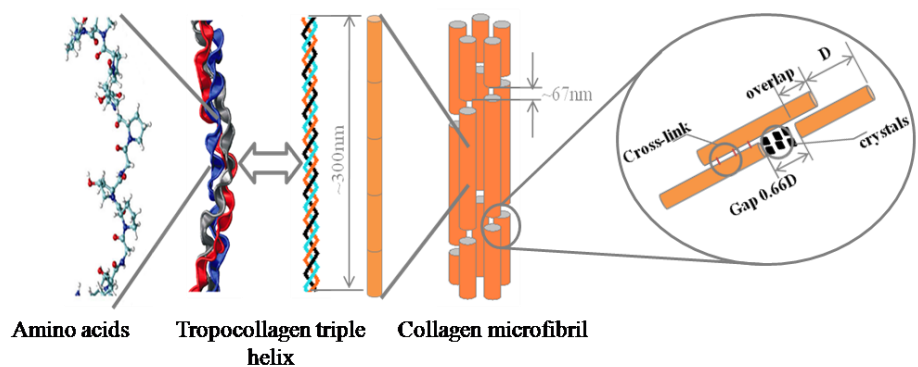

**Figure2**. Schematic illustration of the TC triple helix molecules and the formation of the microfibril

A microfibril is composed of the TC molecule type I linked together by cross-links and mineral phases:

**Tropocollagen molecules:** At the lowest hierarchical level, bone structure is composed of TC molecules which can be viewed as a of rod about 300 nm long and 1.5 nm in diameter, made up of three polypeptide strands, each of which is a left-handed helix (23). The arrangement of the TC molecules comes from the strong chemical bonds (cross-linking) that form between adjacent collagen molecules throughout the collagen bundles (4, 6). Force-strain curve during tensile load of a single TC molecule reported in (4), shows an initial linear elastic regime followed by onset of nonlinear, stiffening behaviour at larger strains beyond approximately 30–35% strain. It has been suggested by (14) that regime I is characterized by uncoiling of the TC molecule and regime II is associated with a larger modulus due to stretching of covalent bonds. In this work the tensile test are conducted in small strain ($\varepsilon$ <2%), that's why the TC molecule are considered with linear elastic behaviour.

**Mineral:** The mineral phase is almost entirely composed of impure hydroxyapatite crystals $Ca_{10}(PO_4)_6(OH)_2$. Those crystals are plate like a shape. The size of the mineral plates varies among different kinds of bones, different animals and even different measurement techniques, e.g., transmission electron microscopy (TEM) and small-angle X-ray scattering (SAXS) (24). A wide range of mineral plate dimensions has been reported in the literature: 15–150 nm in length, 10–80 nm in width and 2–7 nm in thickness, while the distance between the neighboring plates is on same the order as the thickness (25). Hydroxyapatite mineral is stiff and extremely fragile exhibiting elastic isotropic behaviour (26-28)

Mechanical properties of hydroxyapatite and TC molecules crystals reported in the literature are grouped in table 1.

| Table 1. Mechanical properties of phases ( tropocollagen molecules and Hydroxyapatite crystals) | | | |
|---|---|---|---|
| Phases | Young's modulus (GPa) | Poisson's ratio | source |
| Tropocollagen molecules | 2.7 | 0.27 | Sansalone, et al ,2007 (29) |
| | 2.4 | -- | Vesentini, et al 2005 (30) |
| | 0.35-12 | -- | Sun et al 2002 (13) |
| | 1.5 | 0.38 | Wagner and Weiner, 1998 (31) |
| | 2.8-3 | -- | Sasaki and Odajima(1996)(16) |
| Hydroxyapatite crystals | 114 | 0.30 | Wagner and Weiner, 1998(31) |
| | 150 | 0.27 | Cowin ,1989 (32) |
| | 170 | 0.33 | Currey (1969) 33) |

**Cross-links:** Joining two TC molecules. The microfibril structure is stabilized through intermolecular cross-links, formed between telopeptides and adjacent triple helical chains through lysine–lysine covalent. Recently, Uzel et al. (6) proposed a rheological model of cross-links behavior with three regimes: (i) Elastic regime, (ii) delayed response due to the unraveling of the telopeptide and (iii) the friction representing the intermolecular slippage. As the present work focus on the small deformation of microfibril, only the first regime (elastic) was taken in consideration. In order to represent the FE behavior of a cross-link, linear spring elements of tensile stiffness $K_{cr}$ (34) were used:

Where:

$K_{cr}$= 17 *kcal/mol/A$^2$* = 1181.143 e-11*N/nm* (1 *kcal/mol/A* ≈ 69.48 *pN*)

**Geometry FE model:**

A repetitive portion of mineralized collagen microfibril was selected; it is a cylinder-shaped with a diameter of 4 nm. There are some studies suggest that collagen microfibrils have a quasi-hexagonal structure (35). Smith et al. (36) consider that each microfibril consists of exactly five molecules in a generic circular cross section. In the current work, the smith model was adopted.

Fig.3 illustrates the unfolding in plane of chosen repetitive portion with mentioning of dimensions and provision of each phase.

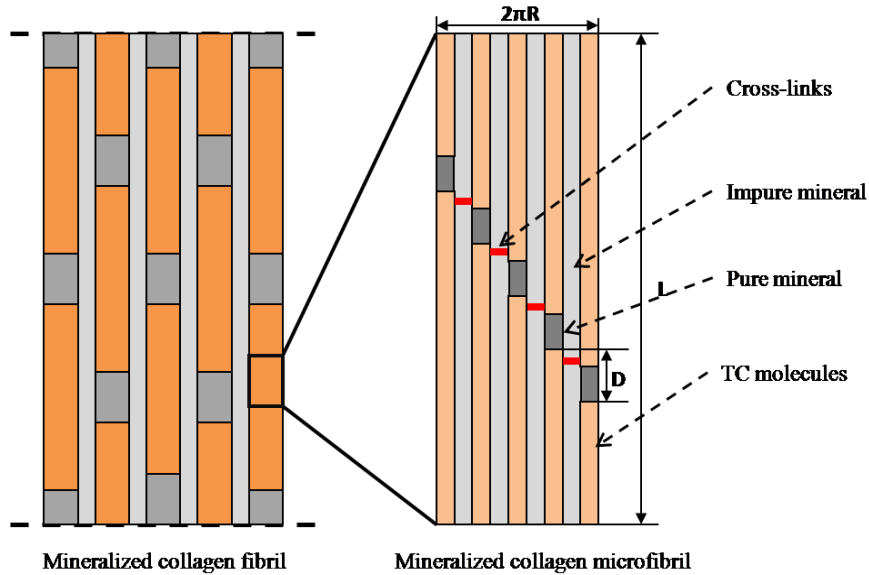

**Figure3**. Illustrated in plans of mineralized collagen fibril and unfolding in plane of repetitive portion of mineralized collagen microfibril structures. The model consists of three phases: (i) TC molecules, (ii) pure mineral: crystalline hydroxyapatite located within gaps created by the TC assembly; (iii) impure mineral: water, crystalline hydroxyapatite and porosity (light grey). Cross-links are connecting the TC molecules.

The proposed model is a periodic repetitive portion with specific geometry properties given in table 2.

| Table 2. geometry properties of proposed model | |
|---|---|
| length of model: L | 340 nm |
| length of tropocollagen molecule: l | 300 nm |
| radius of model: R | ~ 2 nm |
| radius of tropocollagen molecule: r | 0.7 nm |
| discrete gaps (hole zone): G | 40 nm |
| periodicity : D | 67 nm |

A three- dimensional finite element model of mineralized collagen microfibril with symmetric and periodic boundary conditions is considered here, with an array of 5 tropocollagen molecules cross-linked together using springs, the all is put into a mineral matrix. This study focuses on the elastic behaviour of the mineralized collagen microfibril in the small strain regime (ε < 2%). Plasticity and rupture in both phases (TC and mineral) and relative sliding on the interface between two phases was not considered here. The bottom surface of the microfibril was encastred and an uniaxial force (F) along the axis of the collagen molecules was applied to the top surface of the microfibril. Deformation and microfibril elongation is computed by the FE model. Each collagen domain represents a collagen triple helix in a wet environment described as homogeneous elastic phase considering that a single collagen molecule would not exhibit shear deformation. Then each collagen domain deforms only in tension, similar to the situation present in bead models of collagen (4).

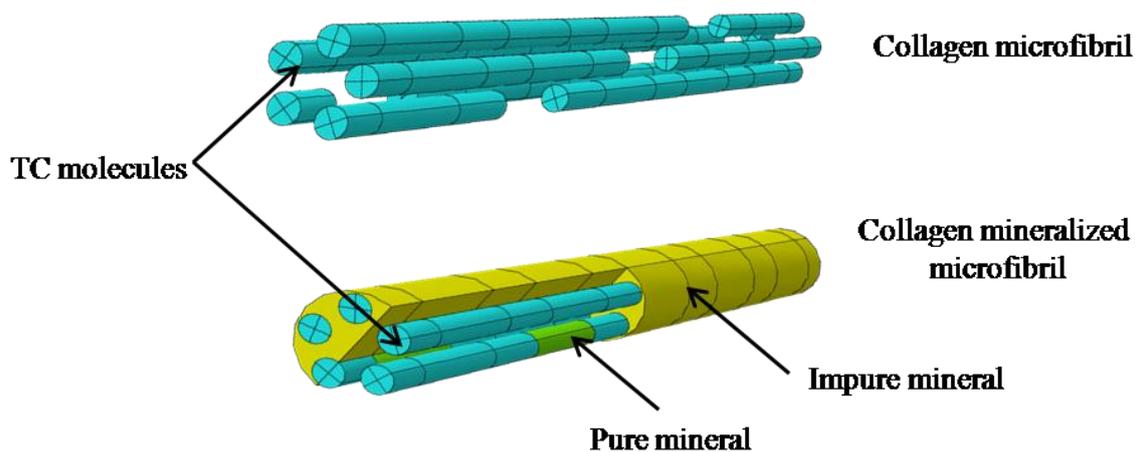

**Figure4.** Proposed 3D FE model of mineralized collagen microfibrils

Both pure and impure portion of mineral have been assumed as homogeneous phase because of these reasons: (i) The aim of this study was to propose a realistic 3D finite element model of mineralized microfibril, present model can be enhanced in future works by including more details and refinements, (ii) To simplify the design of our finite element model and (iii) Due to lack of information dealing with the mechanical properties of the immature phase located at the between the TC molecules in transversal direction. In the finite element model, same mechanical properties are introduced for both phases.

The mineralized collagen microfibril composed of the three phases is considered an elastic material and the constitutive law used is given by:

$$\sigma_{ij} = C_{ijkl}\varepsilon_{kl} \qquad (1)$$

$\sigma_{ij}$ is the Cauchy stress component, $\varepsilon_{kl}$ the linear strains and $C_{ijkl}$ are the components of elasticity tensor

The proposed FE model was coupled to an optimization algorithm based on minimizing to this finite element model to minimize the spring back without having recourse to an experimental database. The developed inverse identification algorithm consists in minimizing the objective function, which is the difference between the predicted force-displacement curves obtained respectively by the equivalent (monophase) model and the 3D simulation using the multiphase model. This method called inverse identification used to calculate equivalent proprieties.

3. **Computational results**

As mentioned above, from the previous literatures, both mechanical and geometrical properties of collagen and mineral have not fixed values because of the nature of the alive human bone tissue. In fact these variations have a significant influence on the properties of the collagen mineralized microfibril in the nano-scale and on the cortical bone in the macro-scale. A parametric study was performed in order to investigate the influence of both geometrical (TC molecule diameter) and mechanical parameters on the mechanical behaviour of the microfibril.

Fig. 5 depicts that the equivalent Young's modulus of the microfibril is increased with increasing of the Young's modulus of the mineral. Fig.6 shows the effect of cross-links number on equivalent Young's modulus. A non linear curve is obtained. It is composed of two parts: (i) from N = 1 to N = 20, the cross-links number has an effect and (ii) a constant value "plateau value". It also shows that when the quantity of collagen increases, the microfibrils ductility increases.

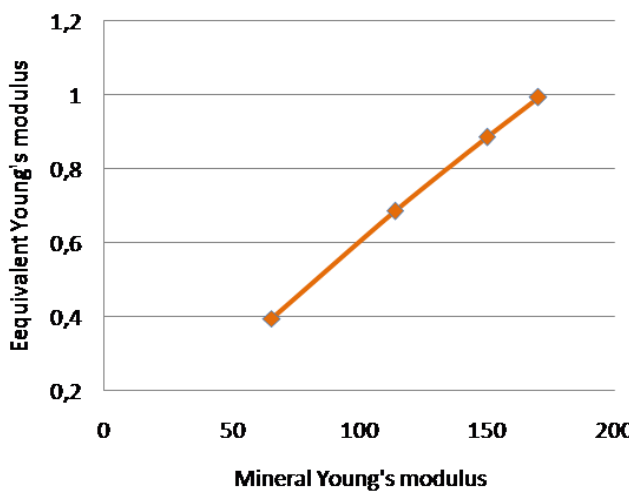
**Figure5**. Equivalent Young's modulus of a collagen microfibril as a function of the Young's modulus of mineral Em at Ec= 2.7GPa and N=1.

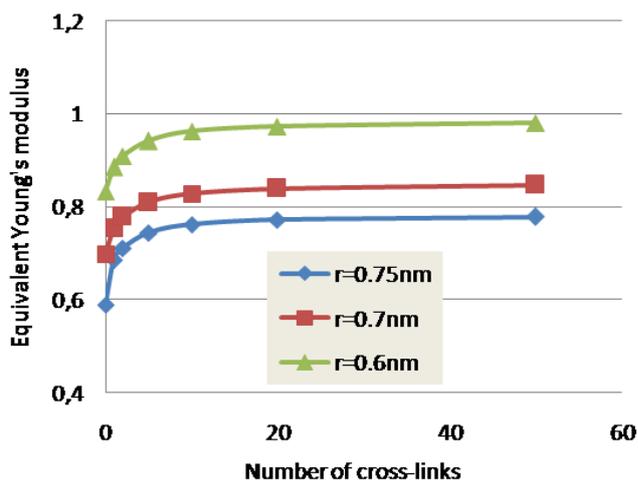
**Figure6**. Equivalent Young's modulus of a collagen microfibril as a function of the cross-links number under varying TC diameter Ec=2.7GPa and Em= 114GPa and N=1.

Fig. 7 shows that if the Young's modulus of the mineral increases, the equivalent Poisson's ratio decreases due to the hardness of the mineral material i.e. as its Young's modulus increases the hardness of the collagen microfibril increases, and its ductility decreases. On other side as the Young's modulus of collagen (soft and elastic material) increases ,the ductility of whole structure increases, which can be expressed by the increase of Poisson's ratio as shown in Fig. 8.

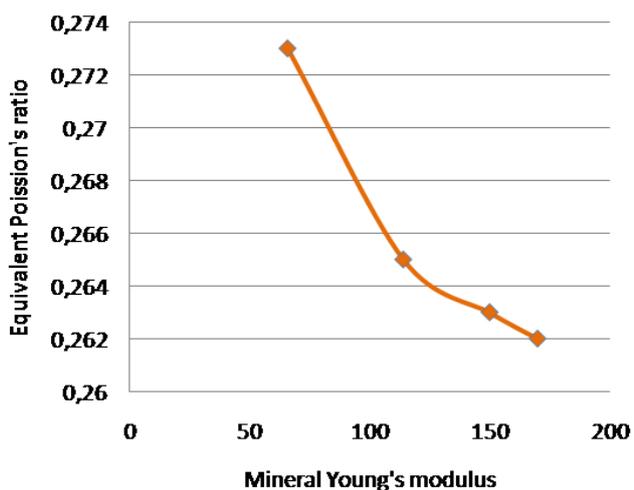
**Figure7**. Passion's ratio of a collagen microfibril as a function of the Young's modulus of mineral Em with Young's modulus of collagen Ec= 2.7GPa and number of cross-links N=1.

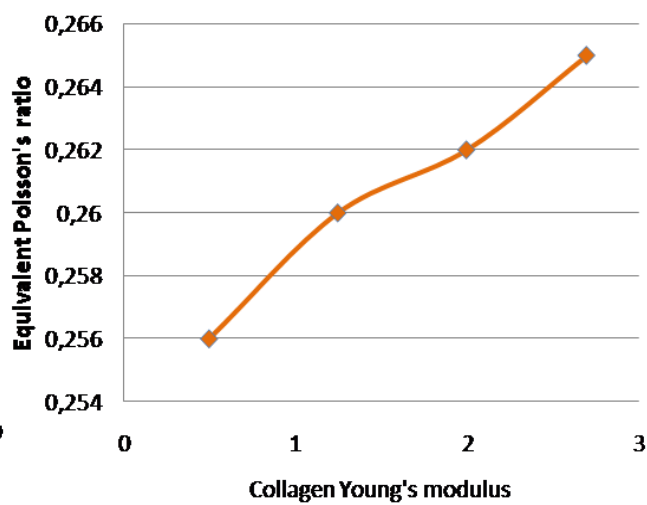
**Figure8**. Passion's ratio of a collagen microfibril as a function of the Young's modulus of collagen Ec with Young's modulus of mineral Em= 114GPa and number of cross-links N=1.

Fig. 9 shows that Von Mises stress increase when the Em increases; it means that the microfibril becomes more rigid and resistant

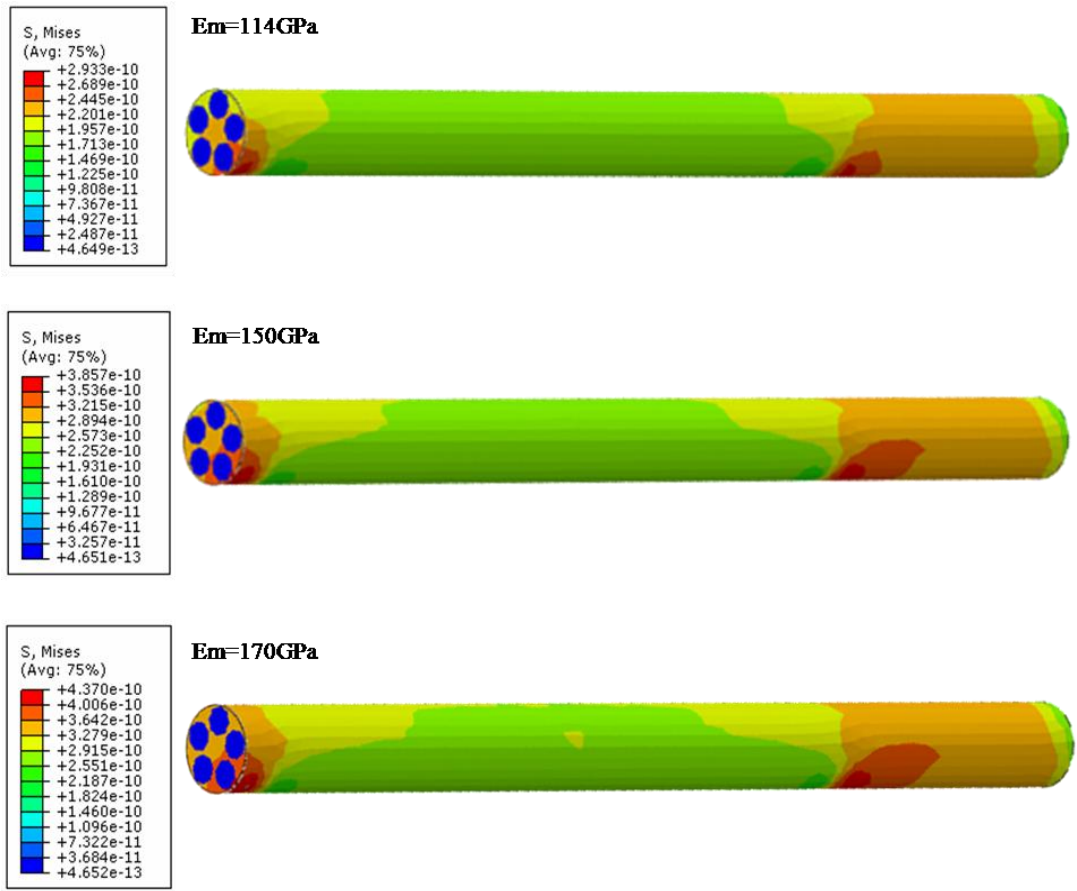

**Figure9.** FE Von Mises stress contours under varying mineral Young's modulus at Ec=2.7GPa and N=0.

### 4. Discussion and Conclusion

Bone is a composite material, where the nanoscale characteristics of individual constituents are important to the overall quality of bone and their mechanical properties (37). On the nanoscale, it is important to determine the influence of the structural properties of individual TC molecules and the single mineralized crystallites within the bone matrix. Physical and mechanical properties of each constituent, such as mineral crystal size and orientation, collagen diameter and orientation, and hardness and elastic modulus of nanoscale structures are all characteristics that contribute to the fragility and strength of mineralized collagen microfibrils.

The collagen network in bone provides resistance against fractures and may be susceptible to change with aging and disease (38). For example, in osteogenesis imperfecta, a disease characterized by decreased material properties and bone fragility, some mutations in the amino-acid sequence of type I collagen can lead to the formation of branched fibers responsible for brittle bone and abnormal mineralization (39). The major role of Collagen, as it is known, is to give bone its ductile properties, so understanding how collagen microfibrils are oriented as well as knowing their geometry is important to study the bone quality (40, 41). The orientation of the collagen within the bone matrix is important to bone stiffness and strength (42, 43). Although the classical convention says that collagen is primarily correlated to the post-yield properties of bone, collagen can have an effect as well on bone's tensile stiffness, and perhaps its strength, although this is less clear (44, 45).

In other side, although collagen may have some influence on the pre-yield properties, the mineral fraction is widely acknowledged to have a greater impact on the pre-yield mechanical properties than collagen does. Collagen content and cross-linking are more closely related to the post-yield properties of bone, and have a profound effect on bone's toughness, ductility, and viscoelasticity (46, 47). There are also studies showing the importance of the mineral phase with its density, mechanical properties, arrangement and orientation of crystal leaves which it is constituted of. Bone mineral density is the gold-standard for assessing bone quantity and diagnosing osteoporosis. Although bone mineral density measurements assess the quantity of bone, the quality of the tissue is an important predictor of fragility (48). Also studies have shown that fluoride treatments for osteoporosis negatively alter crystal structure (49, 50), decreasing elasticity and bone strength and thereby increasing fracture risk, (51, 52). In agreement with these facts, this work confirm that the decreasing of ductility of the collagen microfibrils is caused by the increasing of the Young's modulus of mineral phase, clarified by the curve of the equivalent Passion's ratio as a function of the Young's modulus of mineral phase ( Fig. 7).

The graph Fig.9 shows that the number of cross-link has a most influence on the increase in the equivalent Young's modulus. if increase the number of cross-link increase the bone material will be more stiffness , however when N > 20 Young's modulus does not depend on the cross-link number and it remains at constant value "plateau value", same observation has been also found by J.Buehler (4) in his molecular multi-scale study of collagen fibril, where it is found that the yield stress and fracture stress depend on the cross-link density "β" only when β < 25 and it is interpreted as the plateau value can be explained by a change in molecular deformation mechanism from predominantly shear (for β < 25) to molecular fracture (for β >25) .

Thus, a 3D multiscale study is necessary since the model 2D, although it reflects the mechanical properties of the phases, it doesn't allow us to present microfibrils with their specific arrangements as well the orientation of their TC molecules. Our proposed model combines the two features geometric and mechanical of each phase. The results found confirm the functionality of this 3D finite element model and justify the results found by above theoretical and experimental studies. The individual properties of the collagen fibers and mineral crystals cannot fully explain the mechanical behaviour of the bone matrix. Rather, it is the interaction and orientation of the constituents that could influence bone quality (53), this condition is satisfied by the model being proposed which takes into account the effect of cross-links whose importance maturity to bone's mechanical proprieties stays unclear (54, 55) and the manner in which the molecules of tropocollagen are linked that can have a profound effect on the strength, stiffness and fragility of the tissue.

In this paper we study for the first time the mechanical behaviour of microfibrils using the 3D finite element model and the inverse identification method. The Results found in this study seems logical and consistent among, they prove functionality and reliability of this model which can be also used to investigate many other phenomena and role of same parameters linked to mechanical and biological behaviour of bone in this scale. This work also allows us to understand better this nanoscale and study the upper level scale which is the collagen fibril with the same methods by using the results found in this scale. But before proceeding to the next level to reach cortical bone's one, it is necessary to improve these results by studying the effects related to mechanical characteristic of constituents. With the model being proposed, we try to model microfibrils and touch on the reality, but this model is still limited owing to the specific arrangement of collagen molecules, the orientation of mineral crystals, the complicated distribution of each phase and in the manner of modelling the cross-links

The current work studied the microfibril mechanical behaviour under tensile and compression tests. In tension/compression, the shape of the model is not important since the reaction depends only on the area of the section. The shape play a role during the bending and torsion tests, if the quasi –hexagonal structure of microfibrils is admitted, this model cannot be used for torsional and bending.

**Acknowledgements**

This work has been supported by French National Research Agency (ANR) through TecSan program (Project MoDos, n°ANR-09-TECS-018).